\documentclass[reprint,a4paper,floatfix,superscriptaddress]{revtex4-1}
\usepackage{graphicx}
\usepackage{rotating}
\usepackage{amsfonts,amsmath,latexsym,bbm,dsfont}

\usepackage{color}

\newcommand{\beq}{\begin{eqnarray}}
\newcommand{\eeq}{\end{eqnarray}}

\def\RCS$#1: #2 ${\@namedef{RCS#1}{#2}\typeout{RCS #1: #2}}
\mathchardef\lt="313C \mathchardef\gt="313E
\mathcode`<="4268 \mathcode`>="5269

\renewcommand\vec\mathbf

\begin{document}

\author{Daniel Kats}
\author{Frederick R. Manby}
\affiliation{Centre for Computational Chemistry, School of Chemistry,
  University of Bristol, Bristol BS8 1TS, UK}

\title{The distinguishable cluster approximation}
\begin{abstract}
A new method that accurately describes strongly correlated states and captures dynamical correlation is presented. 
It is derived as a modification
of coupled-cluster theory with single and double excitations (CCSD) through consideration of particle distinguishability between dissociated fragments, whilst retaining the key desirable properties of particle-hole symmetry, size extensivity, invariance to rotations within the occupied and virtual spaces, and exactness for two-electron subsystems.
The resulting method
called the distinguishable cluster approximation, smoothly dissociates difficult cases such as the nitrogen molecule, with the modest $N^6$ computational cost of CCSD. 
Even for molecules near their equilibrium geometries, the new model outperforms CCSD.
It also accurately describes the massively correlated states encountered when dissociating hydrogen lattices, a proxy for the metal-insulator transition, and the fully dissociated system is treated exactly. 
\end{abstract}

\maketitle 
\section{Introduction}
Coupled-cluster theory \cite{cizek} is the most successful method used for the treatment of many-body correlations in quantum chemistry. The wavefunction is represented in exponential form
\begin{equation}
  |\Psi> = e^{\hat T}|0>
  \label{eq:CC}
\end{equation}
where $|0>$ is the Hartree-Fock reference state and $\hat T=\hat T_1+\hat T_2+\cdots$ is a sum
of cluster operators that build correlations into the wavefunction through single, double and higher excitations \cite{shavitt_and_bartlett,crawford_and_schaefer}.
Even when the expansion of $\hat T$ is truncated, coupled-cluster theory retains important properties of the exact solutions to the Schr\"odinger equation, primarily size extensivity, and includes all contributions from all orders of perturbation theory up to the given excitation rank.
Coupled-cluster theory provides a systematically improvable hierarchy of polynomially scaling methods that approach full configuration interaction (or exact diagonalization).

In quantum chemistry, coupled-cluster theory is typically truncated at the doubles level (to give the so-called CCSD method \cite{Purvis:82}), and the effect of triple excitations is handled perturbatively in the CCSD(T) method \cite{Raghavachari:89}. These truncated coupled-cluster theories achieve extremely high accuracy for a very wide range of problems, but fail to describe the strongly correlated states encountered in molecular dissociation or other highly degenerate situations. For example, CCSD famously fails to repair the incorrect physics of a restricted Hartree-Fock treatment of the dissociating N$_2$ molecule, and the perturbative triples correction only makes matters worse. 

An enormous range of approaches have been devised to address strong correlation in quantum chemistry. Those that build on coupled-cluster theory do so by using a more flexible reference function \cite{mukherjee_correlation_1975,kohn_state-specific_2013}; by symmetry breaking \cite{krylov_size-consistent_2001}; by including leading terms from variational coupled-cluster theory \cite{robinson_approximate_2011}; through renormalization \cite{kowalski_method_2000}; or through combining valence-bond theory and coupled-cluster \cite{small:114103}.

One could assume that CCSD is the most accurate theory possible that includes only single and double excitations from a single reference determinant, because it includes all possible diagrammatic contributions to the correlation energy within that constraint. But consider  some quantity in the full theory $x=y+z$ that is approximated in CCSD as $x_\text{SD}=y_\text{SD}+z_\text{SD}$. Now if $z\approx 0$ (for example through cancellation of $z_\text{SD}$ with higher-order terms), it could be better to exclude the $z_\text{SD}$ terms altogether and use the approximation $x_\text{SD}=y_\text{SD}$ instead.

In this spirit, there is a growing body of evidence that improved doubles-based schemes can be produced by leaving certain terms out of the amplitude equations, like various versions of the coupled-electron pair approximation \cite{Meyer:71, Neese:09}, the $n$CC hierarchy \cite{bartlett_addition_2006} and pCCSD \cite{huntington_pccsd:_2010}. 

\section{Antisymmetry of the wavefunction}
The traditional description of antisymmetry in quantum mechanics states that 
$$
  \Psi(\vec r_1,\vec r_2) = -\Psi(\vec r_2,\vec r_1)
$$
noting that this (or its bosonic counterpart) is an unavoidable consequence of particle indistinguishability. The soundness of this statement has been called into question, because the notion that the classical configuration $(\vec r_1,\vec r_2)$ is different from $(\vec r_2,\vec r_1)$ distinguishes the particles from the outset \cite{m73}. 
This subtlety appears to have motivated the analysis by Leinaas and Myrheim of the classical configuration space of indistinguishable particles, famously leading to the prediction of anyon statistics \cite{lm77}. 

A concrete meaning can be attached to the concept of particle interchange, as being the result of closed paths in the classical configuration space for indistinguishable particles, in which the points $(\vec r_1,\vec r_2)$ and $(\vec r_2,\vec r_1)$ are identified \cite{lm77}. For the present work, the important result from Leinaas and Myrheim is that although the topology of the configuration space for multiple identical particles is globally different from the Euclidean product space, it is locally isometric in regions where particles are not close.
To express it another way, there are no physical consequences of the indistinguishability of particles during processes in which the particles do not become close.

The problems with truncated coupled-cluster theories are encountered when molecules are fragmented into separate pieces. All possible exchange processes between the fragments are considered, because antisymmetry is cemented into the structure of the theory right from the beginning, through the use of second quantization. In exact theory it makes no difference whether these exchange processes are considered or not (but of course the exchange processes \emph{within} the fragments must be treated exactly). Here we speculate  that an incomplete treatment of exchange processes, whose effect must ultimately cancel out in an exact treatment, could be the source of the pathological failure to describe dissociation in coupled-cluster theory.

\section{Theory}
We have motivated removal of exchange terms between isolated fragments. There is no systematic way to do so in an orbital-invariant theory, so as a substitute we instead investigate removal of exchange terms between the 2-particle clusters formed through the application of the $\hat T_2$ cluster operator, whilst retaining all terms arising from particle indistinguishability \emph{within} clusters.

\begin{figure}
\begin{center}
\includegraphics[height=10ex]{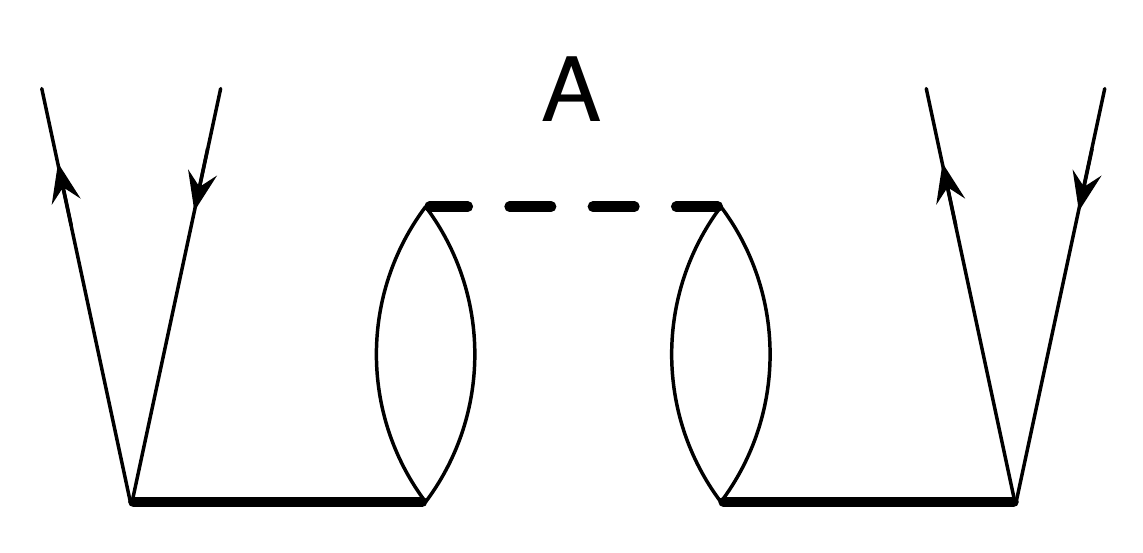}\qquad
\includegraphics[height=10ex]{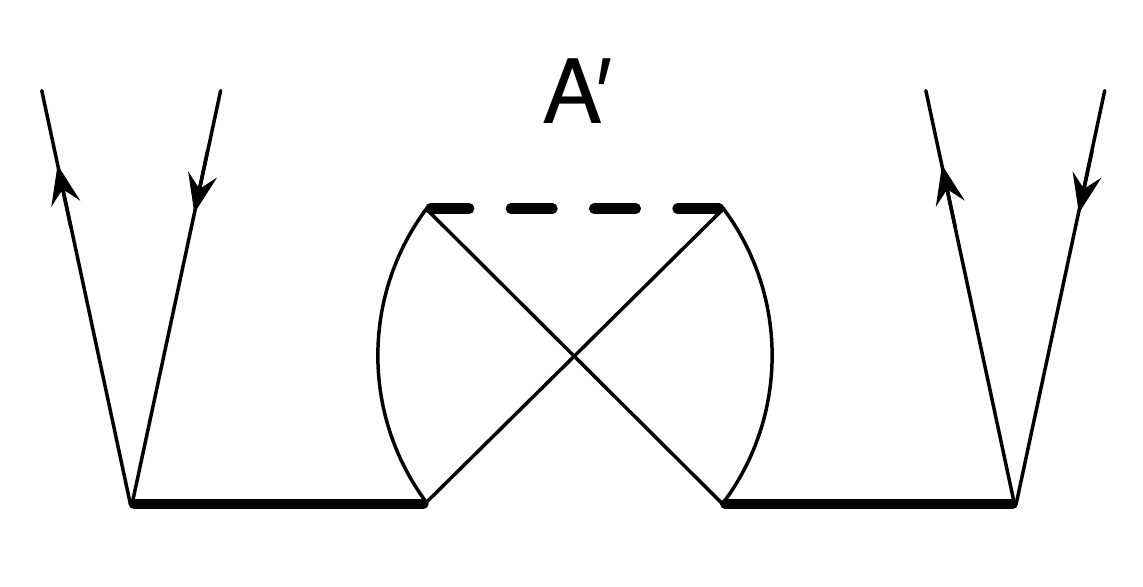}\\[2ex]
\includegraphics[height=10ex]{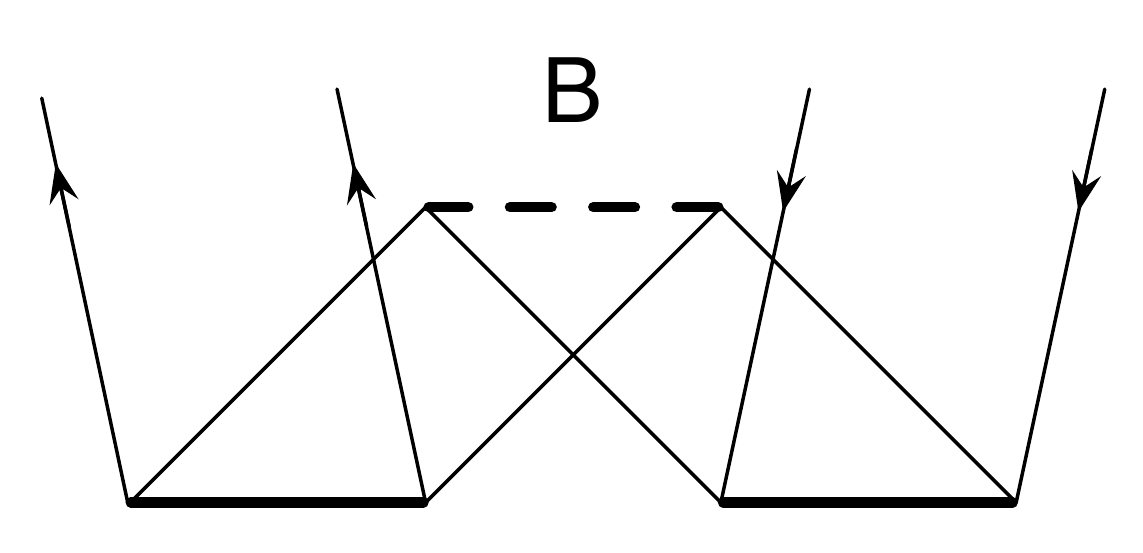}\hfill
\includegraphics[height=10ex]{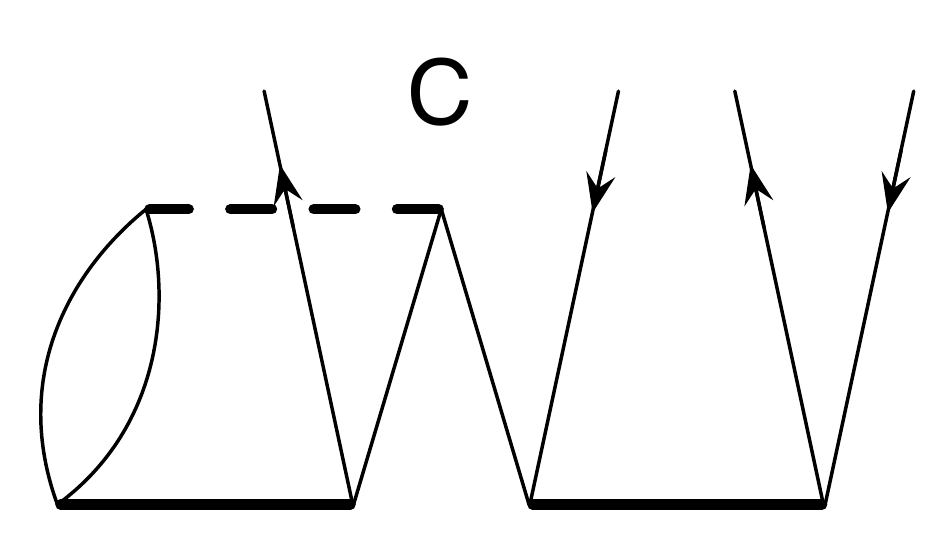}\hfill
\includegraphics[height=10ex]{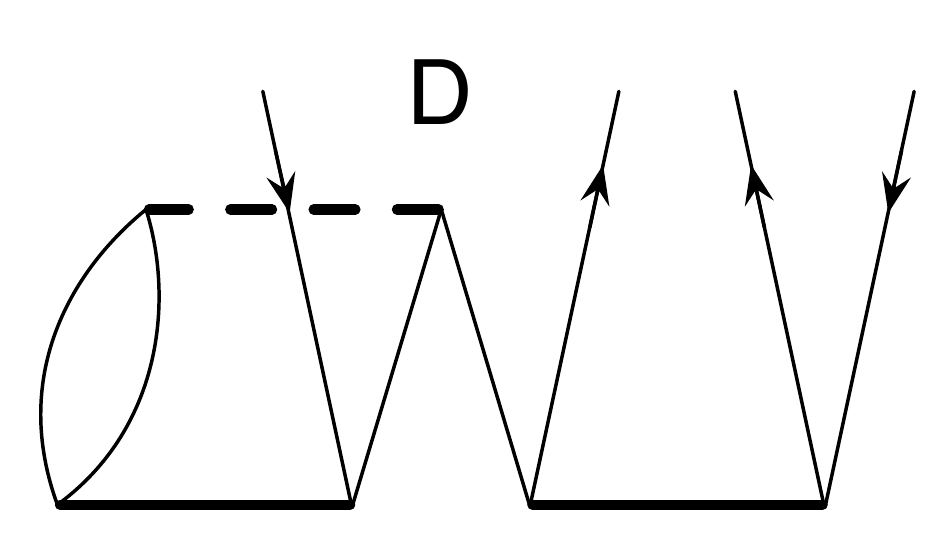}
\end{center}
\caption{Nonantisymmetrized Goldstone diagrams contributing to the CCD amplitude equation that are quadratic in the amplitudes. The solid interaction lines denote a $T_2$ amplitude; the dashed lines a Coulombic interaction from the Hamiltonian; and the thin lines particles (downward arrows) or holes (upward arrows) in the spinorbital representation.}
\label{fig:quadratic}
\end{figure}

We therefore turn our attention to the terms in the amplitude equation that couple together two $T_2$ amplitudes, shown in Figure~\ref{fig:quadratic} in the form of nonantisymmetrized Goldstone diagrams. Contrary to the usual convention, our interaction lines correspond to bare electron repulsion integrals, rather than antisymmetrized integrals.

Consider diagrams \textsf A and \textsf A$'$. In \textsf A, particle-hole pairs formed as a result of two separate double excitation processes interact through the Hamiltonian. In \textsf A$'$ there is an additional process in which the particles (or equivalently holes) from each double excitation are exchanged. In an infinitely separated system the only physically relevant double excitation processes are those where the two particles and two holes are associated with a single fragment. But the very cases where CCSD fails to describe dissociation are those where the restricted Hartree-Fock orbitals cannot be localized on to the fragments. 

We hypothesize that the approximate description of physically irrelevant exchange processes could be the source of the poor behaviour of CCSD for dissociation. 
Removal of diagram \textsf A$'$
might well improve the description in the long range --- in fact it does --- but 
it leads to parity violation because contributions of the form
$$
t^{\cdots i}_{\cdots a}<ij|aa>t^{j\cdots}_{a\cdots}\;\propto
\vcenter{\hbox{\includegraphics[height=7.14ex]{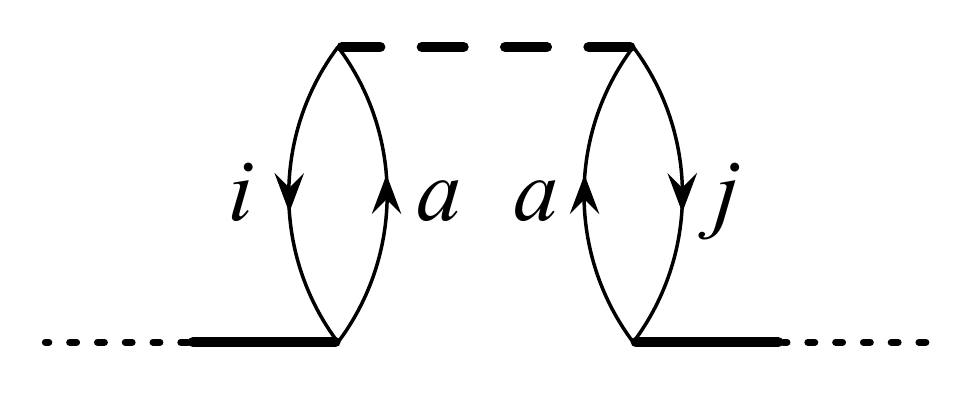}}}
$$
are not cancelled through the corresponding diagonal component of diagram \textsf A$'$.
Nevertheless \emph{within} clusters antisymmetry is handled correctly (the amplitudes have the correct antisymmetry); it is only between clusters that parity violation is permitted.

We now proceed by restoring the other desirable properties of CCD in the new theory. The quadratic portion of the CCD $T_2$ amplitude equation reads
$\mathsf A+\mathsf A'+\mathsf B+\mathsf C+\mathsf D$. The amplitude equation can be modified without compromising exactness for two-electron systems by adding
$\alpha'(2\mathsf A'+\mathsf C)+\beta(2\mathsf B+\mathsf D)$ for any parameters $\alpha'$ and $\beta$. Diagram $\mathsf A'$ is removed by setting
$\alpha'=-1/2$, and particle-hole symmetry then requires $\beta=-1/2$, leading to the modified quadratic contribution
$\mathsf A+\mathsf C/2+\mathsf D/2$.
This defines the distinguishable cluster approximation with double excitations (DCD).

The diagrams in Figure~\ref{fig:quadratic} can be interpreted as a Coulombic interaction between fluctuations in separate clusters (\textsf A), exchange interactions between clusters (\textsf A$'$ and \textsf B), and scattering of a particle (or hole) of one cluster with a Fock-like potential arising from fluctuations in the other (\textsf C and \textsf D). DCD can be viewed as a method in which each two-particle cluster is treated exactly in an embedding produced by other fluctuations in the system, but with neglect of exchange processes between the two-particle cluster and the environment.

To avoid the complications introduced by a full consideration of single excitations we use a Brueckner formulation \cite{nesbet_brueckners_1958,cizek:80,bartlett:80} that rotates the occupied and virtual spaces such that the $T_1$ amplitudes vanish. The BDCD program is implemented as a variation of the BCCD program \cite{hampel_comparison_1992} in Molpro \cite{MOLPRO_brief,MOLPRO-WIREs}.

\section{Results}

\begin{figure}
\begin{center}
\includegraphics[width=\linewidth]{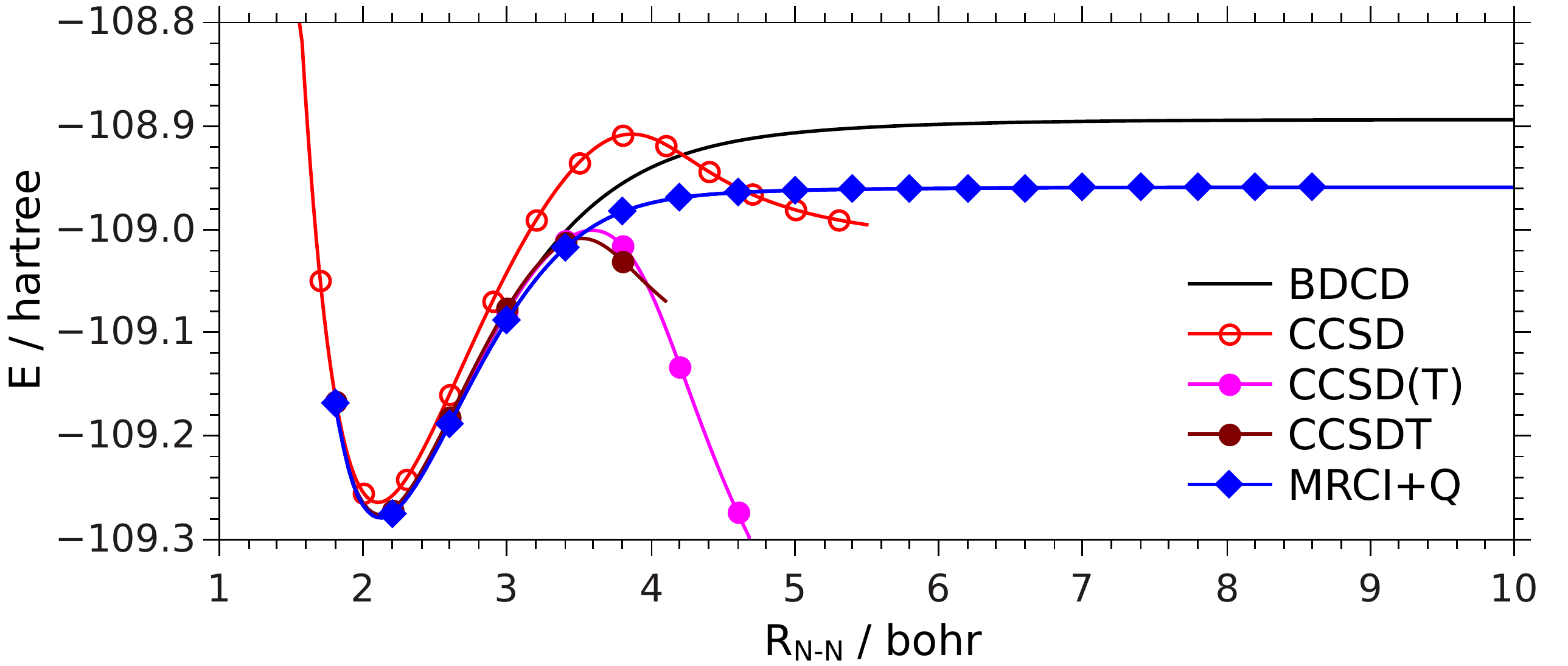}
\includegraphics[width=\linewidth]{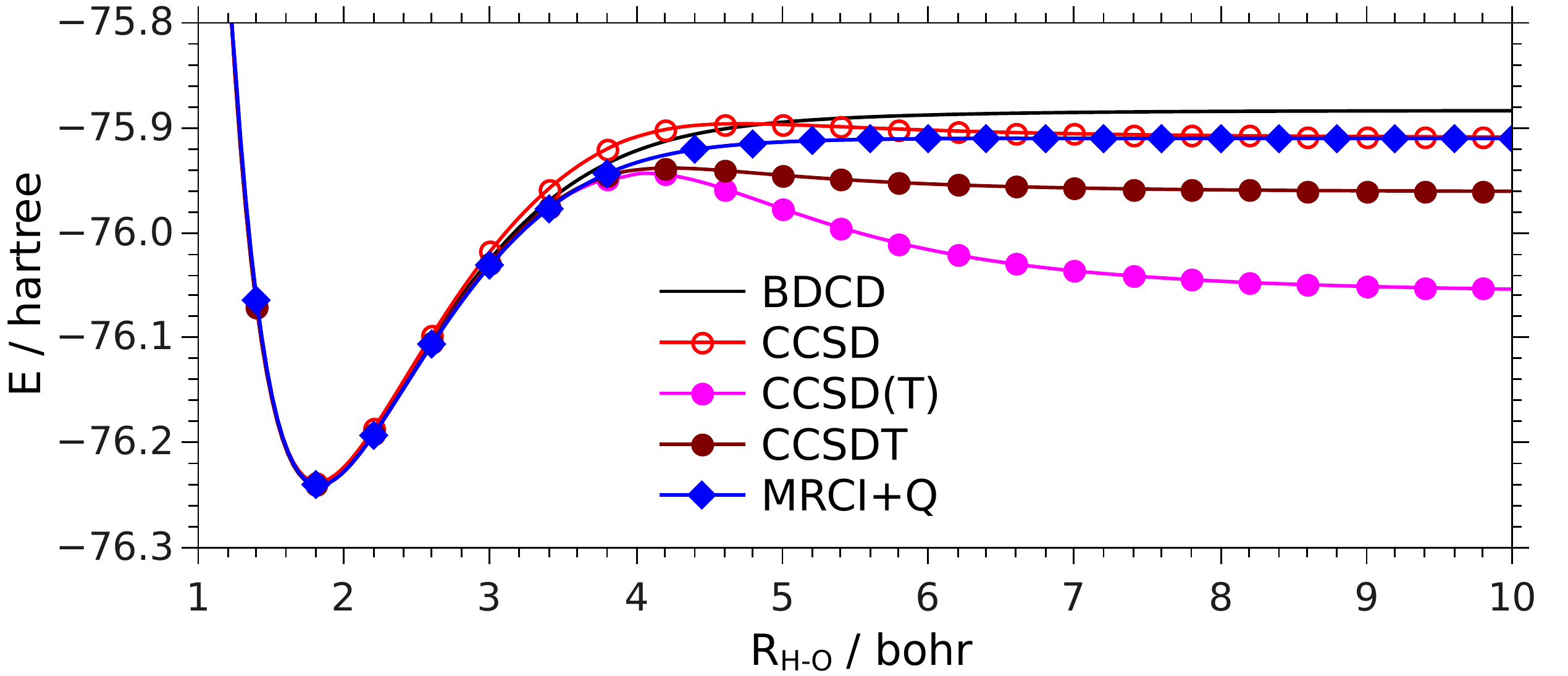}
\end{center}
\caption{Potential energy curves for N$_2$ dissociation (top); symmetric double dissociation of water (bottom). Shown are BDCD and, for comparison, a variety of alternative methods. In each case the hierarchy CCSD, CCSD(T), CCSDT breaks down in the strongly correlated region, but BDCD is systematically smooth and well behaved.}
\label{fig:dissoc}
\end{figure}

We first consider the chemically relevant situation where a low-cost correlation method that can handle strongly correlated states is of immediate importance: molecular dissociation. Calculations were performed using the cc-pVDZ gaussian basis set and were based on a restricted Hartree-Fock reference. In Figure~\ref{fig:dissoc} the BDCD results for N$_2$ and H$_2$O dissociation are compared with other coupled cluster calculations and benchmark data from the Davidson-corrected internally contracted multireference configuration interaction method \cite{WK88} (denoted MRCI+Q). Whereas CCSD, CCSD(T) and CCSDT show increasingly poor behaviour in the strongly-correlated region, BDCD smoothly and systematically dissociates the molecules, resulting in the physically correct description of zero net force between the dissociated fragments. This is despite the fundamentally flawed description in the closed-shell restricted reference state.

The modification to improve description of a fragmented system could well have made things worse in the bonded region, but in fact this turns out not to be the case. On the contrary, the shape of the curve around the minimum is considerably better in BDCD than in BCCD (or CCSD). For N$_2$, the difference in harmonic frequency relative to MRCI+Q is 99\;cm$^{-1}$ for BCCD but only 13\;cm$^{-1}$ for BDCD, and the discrepancy in the equilibrium bond length falls from 0.90\;pm to 0.16\;pm.
Moreover reaction energies are well described by BDCD. For a test set of small-molecule reactions \cite{knizia_simplified_2009} calculated using the aug-cc-pVTZ basis set, the root mean squared deviation compared to CCSD(T) is reduced from 7.6\ kJ/mol for BCCD to 5.0\ kJ/mol for BDCD.

We next consider hydrogen systems, which are important model problems that connect the challenges of strong correlation in the quantum chemistry of dissociating molecules with those in the description of the metal-insulator transition in condensed-matter physics. 

First we consider H$_4$ using the cc-pVQZ basis set. The four atoms sit on the circumference of a circle of radius 1.738\;\AA, with opposite pairs connected by diameters subtending an angle $\theta$ at the centre \cite{van_voorhis_benchmark_2000}.
For $\theta\lt90^\circ$ the electronic structure is dominated by a reference state which pairs the hydrogen atoms into molecules one way, and for $\theta\gt90^\circ$ the other. The MRCI+Q data can be regarded as essentially exact, and the energy goes through a maximum at $\theta=90^\circ$ with zero derivative. Neither CCSD nor BDCD reproduce this feature exactly, both producing two different energies depending on the choice of reference state, but in BDCD the maximum is much more accurately captured, with the energy at $\theta=90^\circ$ being reproduced almost exactly.

Next we consider dissociation of the H$_6$ ring and linear H$_{50}$ using the STO-6G basis set. In the former case the exact result (from full configuration interaction) is available. In the latter case exact diagonalization is unfeasible --- the Hilbert space has over $10^{28}$ dimensions --- but highly accurate results \cite{hachmann_multireference_2006} can be found using the density matrix renormalization group (DMRG) approach \cite{white_density_1992,chan_highly_2002,chan_density_2011}, owing to the linear structure. For H$_6$ it can be seen that BCCD (approximately equivalent to CCSD) rapidly diverges from the exact result, whereas BDCD closely follows the exact curve at all distances, reproducing it exactly in the long range. For H$_{50}$ agreement with DMRG is also spectacular, and to emphasize the extent of the error that BDCD has to correct we also present the reference energy from restricted Hartree-Fock theory.

\begin{figure}
\begin{center}
\includegraphics[width=\linewidth]{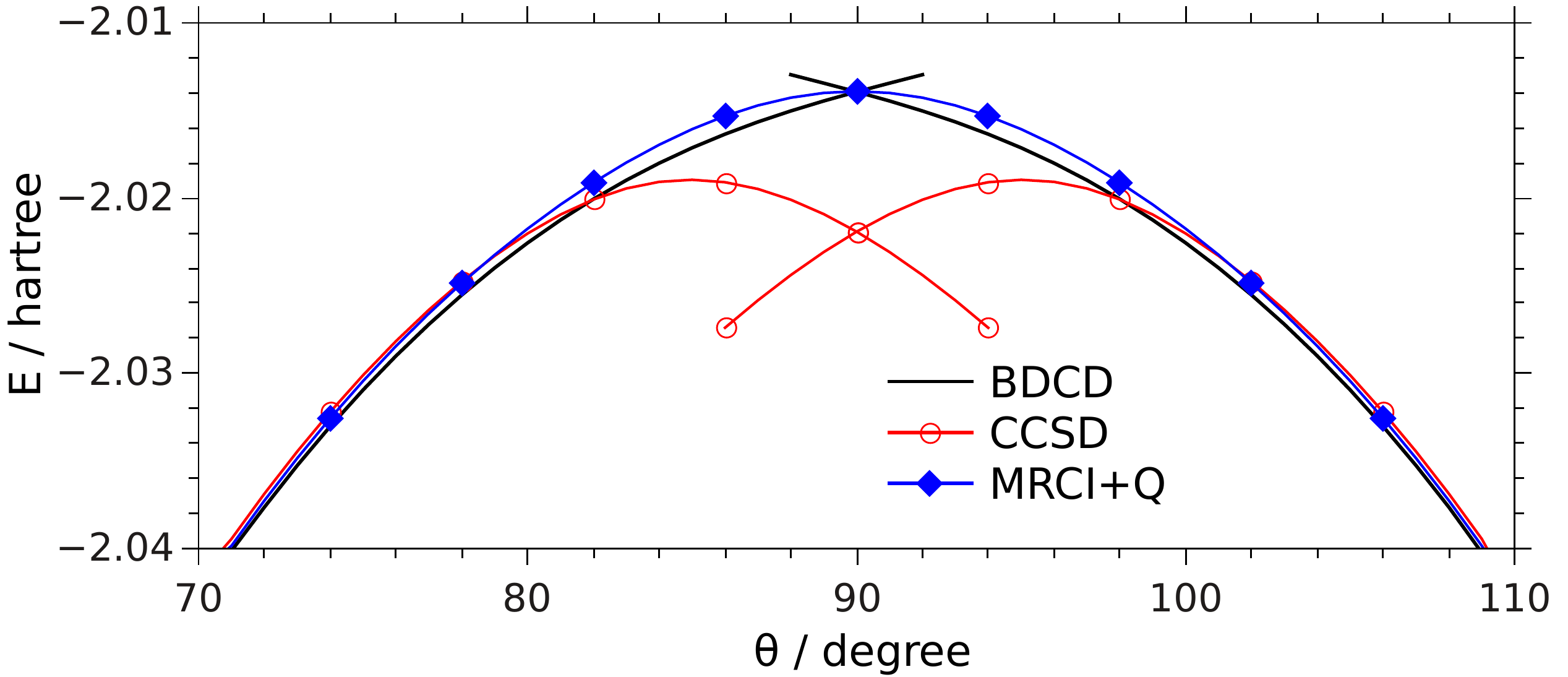}\\
\includegraphics[width=\linewidth]{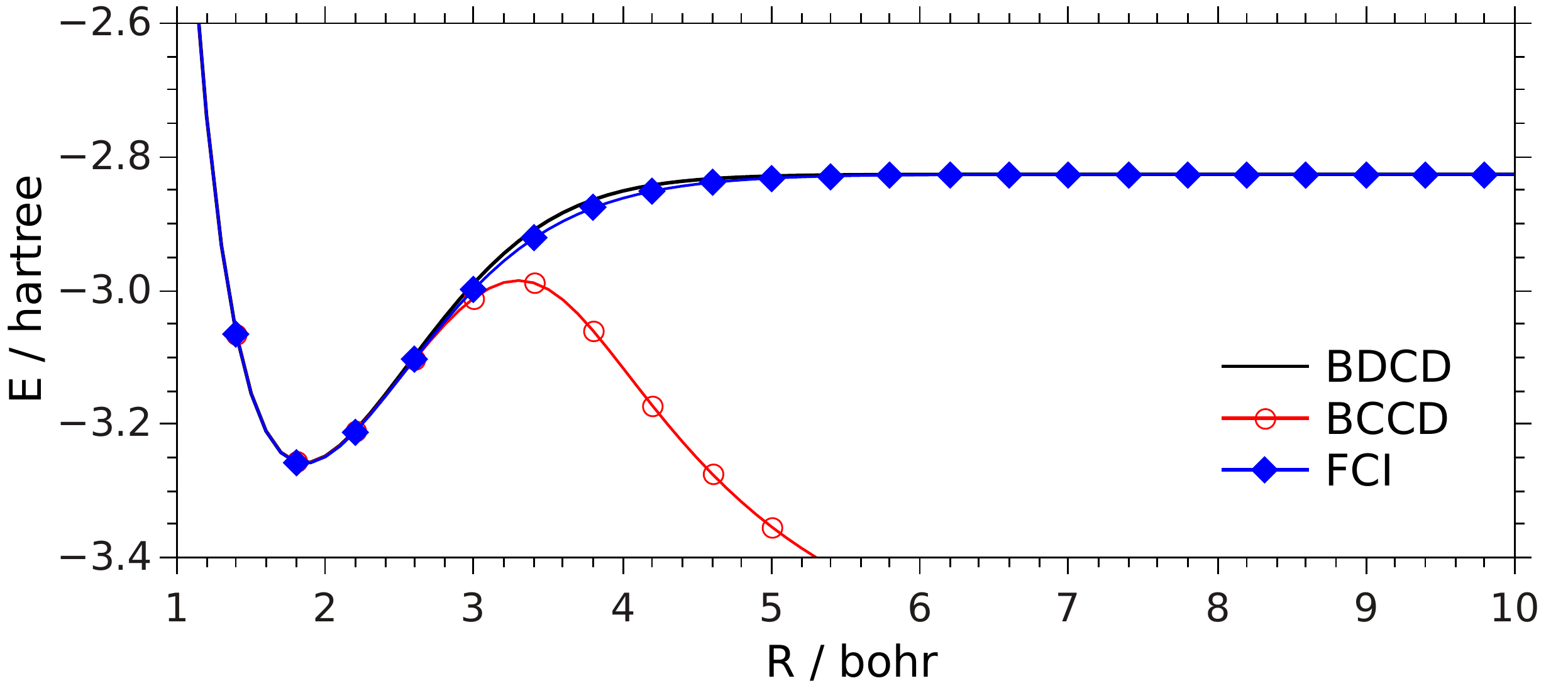}\\
\includegraphics[width=\linewidth]{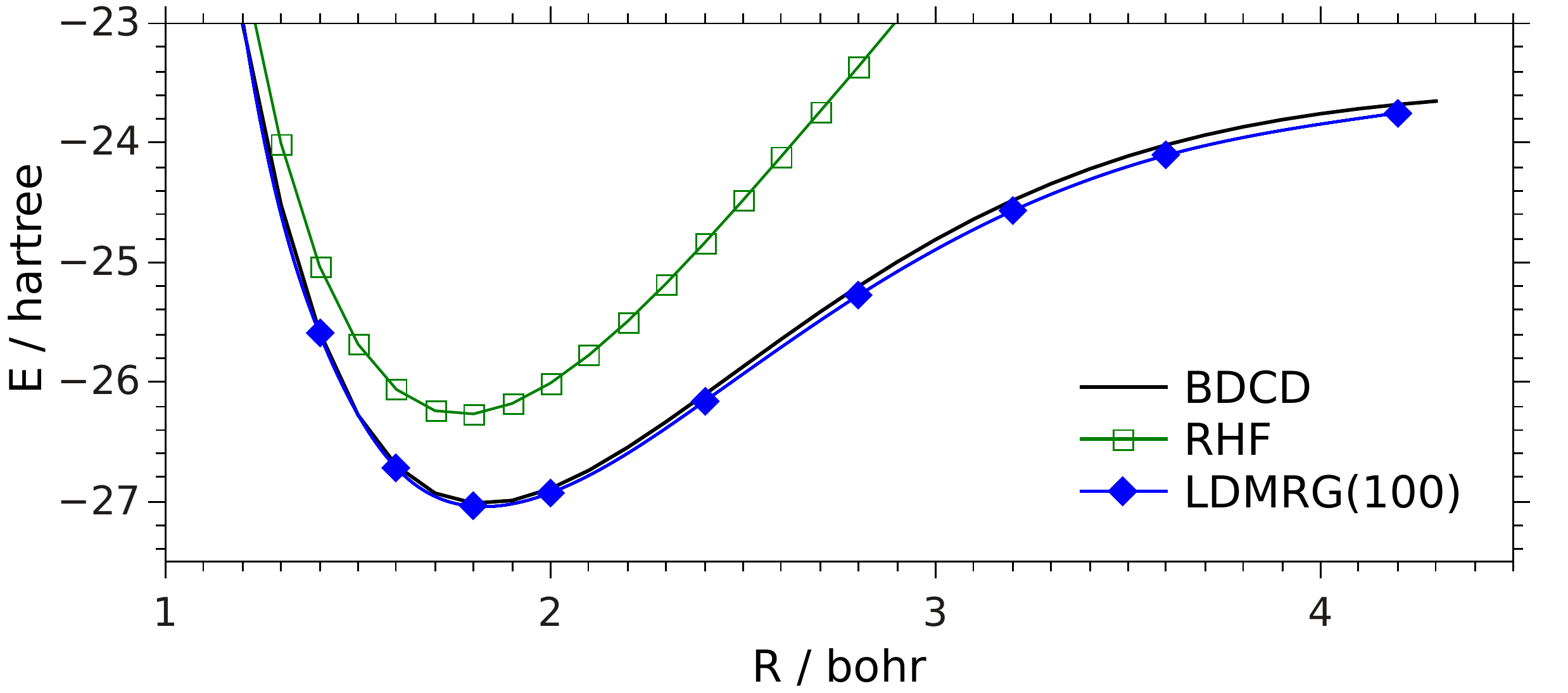}
\end{center}
\caption{Potential energy curves for hydrogen systems. Top: H$_4$ system with geometry from Ref.~\onlinecite{van_voorhis_benchmark_2000} (see text). Middle: uniform dissociation of hexagonal H$_6$, in which $R$ is the distance of each hydrogen atom from the centre. Bottom: uniform dissociation of H$_{50}$ chain, with comparison to DMRG calculations from Ref.~\onlinecite{hachmann_multireference_2006}.}
\label{fig:hsystems}
\end{figure}

\begin{figure*}
\begin{center}
\includegraphics[width=.47\linewidth]{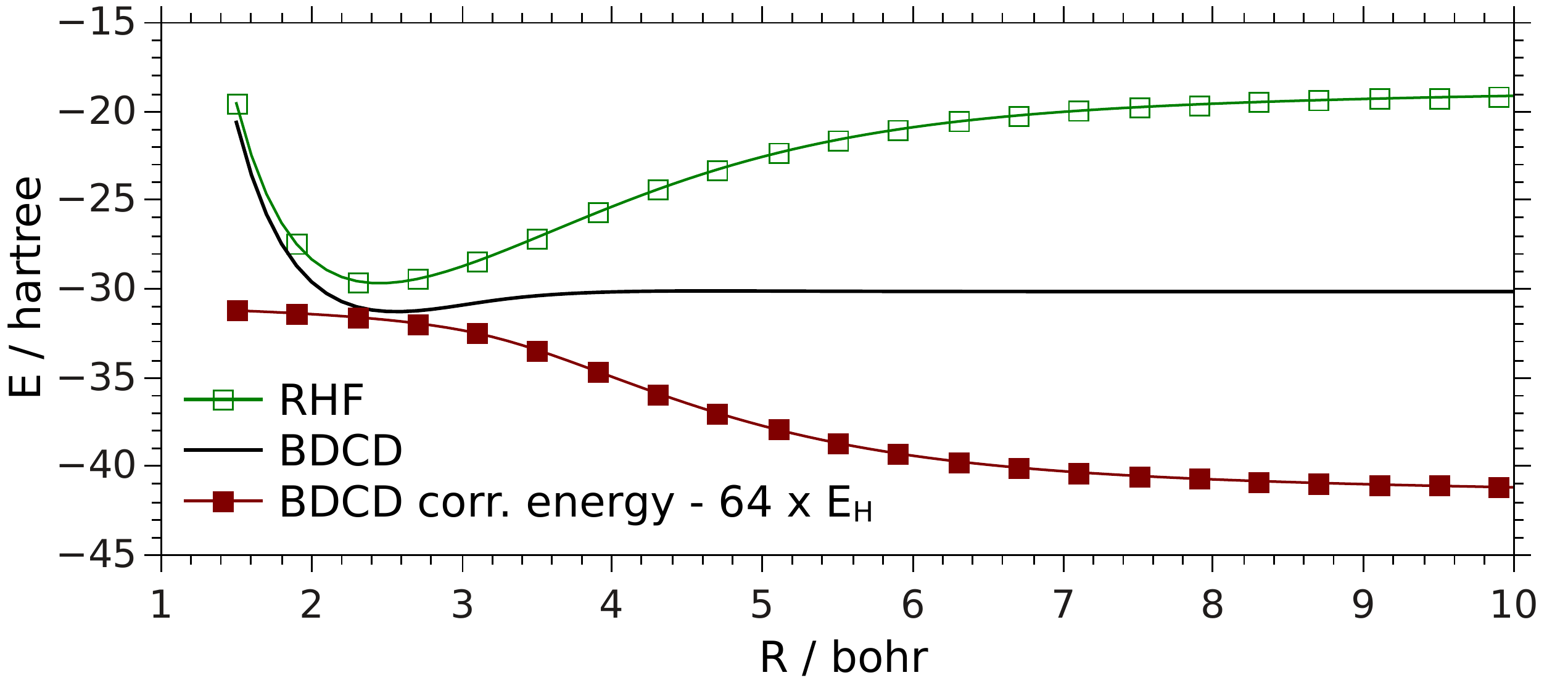}
\includegraphics[width=.47\linewidth]{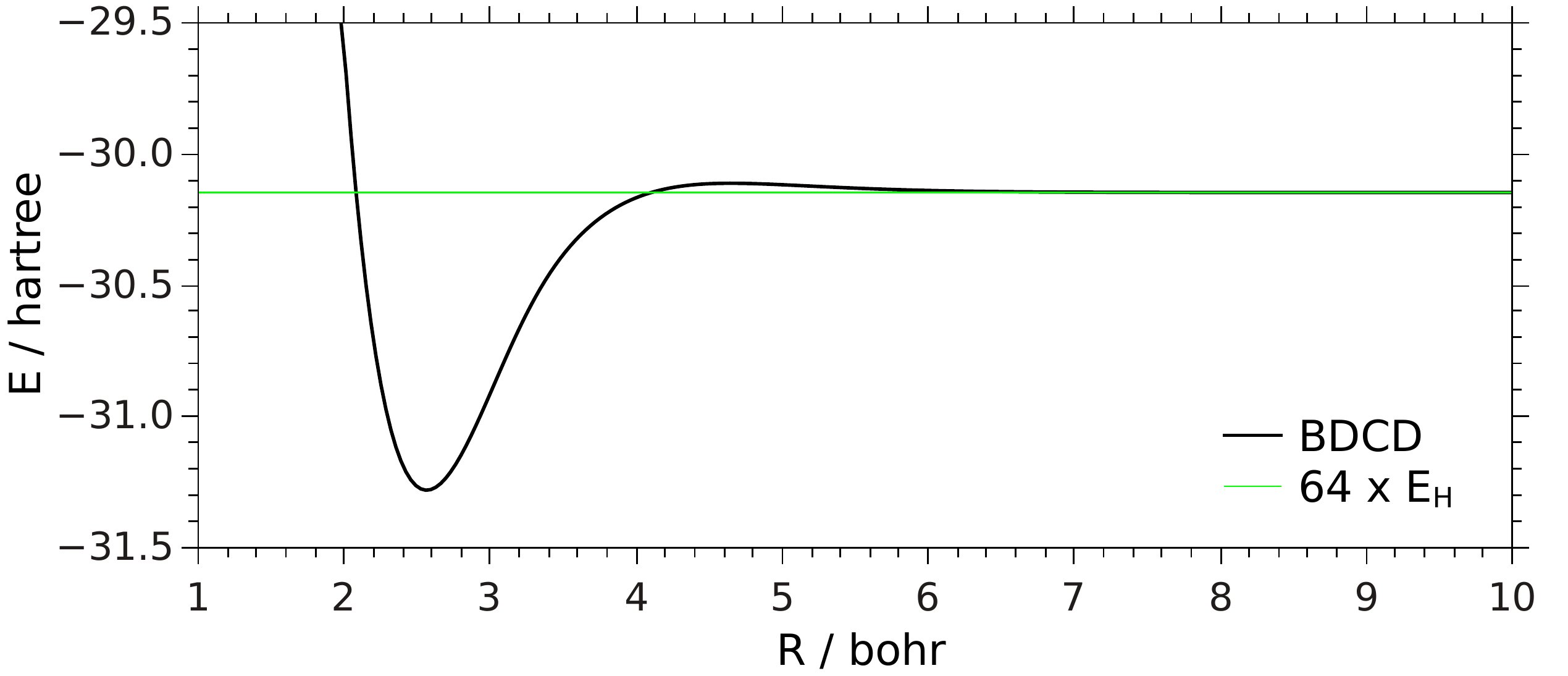}
\end{center}
\caption{Uniform dissociation of $4\times4\times4$ cube of hydrogen atoms, a model for incipient strong correlation in the metal-insulator transition. Left: illustration of the cancellation between the unphysical behaviour of RHF and the correlation energy of BDCD. Right: zoomed view illustrating dissociation to the correct asymptotic value, and a small non-physical feature in the potential.}
\label{fig:h64}
\end{figure*}

A significant challenge for many-body theory is the description of the onset of strong correlation during the metal-insulator transition; this has often been investigated using the model system of cubic lattices of hydrogen atoms \cite{tsuchimochi_strong_2009,sinitskiy_strong_2010}. In the equilibrium geometry the electronic structure is relatively benign, but as the nearest-neighbour distance is increased enormous degeneracy is introduced until, at dissociation, the wavefunction is composed of around $10^{18}$ equally weighted determinants, from a Hilbert space of $10^{36}$ dimensions.

For this system, very few approximations work even qualitatively; the restricted Hartree-Fock energy rises rapidly to spuriously high values; perturbative methods produce infinities; and coupled-cluster calculations appear impossible to converge. BDCD provides a converged energy at all separations, shown in Figure~\ref{fig:h64}, smoothly dissociating to exactly 64 times the energy of the single hydrogen atom (in the STO-6G basis set).

\section{Conclusion}
Coupled-cluster theory is an exceptionally powerful framework for describing many-body correlation effects with polynomial cost. But its structure --- forming a wavefunction through excitations from a single reference Slater determinant --- seems to preclude treatment of strongly correlated states, in which potentially vast numbers of determinants contribute with comparable weights.

Motivated by speculations about the role of antisymmetry in the description of dissociated fragments, we have argued for a simple modification of the CCD amplitude equation to provide a theory that neglects certain exchange terms that must cancel out in an exact description of the dissociated system. By doing so, we also threw out terms important for the description of correlations within the fragments themselves, or in a non-dissociated molecule; but it transpired that all of the desirable properties of CCD could be restored by other modifications. Thus BDCD has $N^6$ cost, is extensive and invariant to orbital transformations; is exact for two-electron systems, and treats particles and holes symmetrically.

The resulting method does indeed describe dissociation of molecules. But perhaps more surprisingly it also improves the energetics in the bonding region, and improve energy differences such as reaction energies. Most astonishing of all, though, is the fact that BDCD smoothly dissociates hydrogen lattices producing exactly the correct energy at infinite separation, a feat that should apparently only be possible using a wavefunction of immense and computationally intractable complexity.

Our attention is now focused on the derivation of this theory in a more rigorous theoretical framework, with the aim of treating singles, doubles and higher excitations in one systematically improvable hierarchy of distinguishable cluster theories.

\begin{acknowledgements}
The authors are grateful to Dr David Tew and Dr Denis Usvyat for helpful discussions during the writing of this manuscript, and DK thanks the Deutsche Forschungsgemeinschaft for funding (Research Fellowship Ka 3326/1). We are grateful to Professor Hans-Joachim Werner for permission to use a modified version of the BCCD code in Molpro for the present calculations.
\end{acknowledgements}


\begin{thebibliography}{32}%
\makeatletter
\providecommand \@ifxundefined [1]{%
 \@ifx{#1\undefined}
}%
\providecommand \@ifnum [1]{%
 \ifnum #1\expandafter \@firstoftwo
 \else \expandafter \@secondoftwo
 \fi
}%
\providecommand \@ifx [1]{%
 \ifx #1\expandafter \@firstoftwo
 \else \expandafter \@secondoftwo
 \fi
}%
\providecommand \natexlab [1]{#1}%
\providecommand \enquote  [1]{``#1''}%
\providecommand \bibnamefont  [1]{#1}%
\providecommand \bibfnamefont [1]{#1}%
\providecommand \citenamefont [1]{#1}%
\providecommand \href@noop [0]{\@secondoftwo}%
\providecommand \href [0]{\begingroup \@sanitize@url \@href}%
\providecommand \@href[1]{\@@startlink{#1}\@@href}%
\providecommand \@@href[1]{\endgroup#1\@@endlink}%
\providecommand \@sanitize@url [0]{\catcode `\\12\catcode `\$12\catcode
  `\&12\catcode `\#12\catcode `\^12\catcode `\_12\catcode `\%12\relax}%
\providecommand \@@startlink[1]{}%
\providecommand \@@endlink[0]{}%
\providecommand \url  [0]{\begingroup\@sanitize@url \@url }%
\providecommand \@url [1]{\endgroup\@href {#1}{\urlprefix }}%
\providecommand \urlprefix  [0]{URL }%
\providecommand \Eprint [0]{\href }%
\providecommand \doibase [0]{http://dx.doi.org/}%
\providecommand \selectlanguage [0]{\@gobble}%
\providecommand \bibinfo  [0]{\@secondoftwo}%
\providecommand \bibfield  [0]{\@secondoftwo}%
\providecommand \translation [1]{[#1]}%
\providecommand \BibitemOpen [0]{}%
\providecommand \bibitemStop [0]{}%
\providecommand \bibitemNoStop [0]{.\EOS\space}%
\providecommand \EOS [0]{\spacefactor3000\relax}%
\providecommand \BibitemShut  [1]{\csname bibitem#1\endcsname}%
\let\auto@bib@innerbib\@empty
\bibitem [{\citenamefont {\v{C}\'{\i}\v{z}ek}(1966)}]{cizek}%
  \BibitemOpen
  \bibfield  {author} {\bibinfo {author} {\bibfnamefont {J.}~\bibnamefont
  {\v{C}\'{\i}\v{z}ek}},\ }\href {\doibase 10.1063/1.1727484} {\bibfield
  {journal} {\bibinfo  {journal} {J. Chem. Phys.}\ }\textbf {\bibinfo {volume}
  {45}},\ \bibinfo {pages} {4256} (\bibinfo {year} {1966})}\BibitemShut
  {NoStop}%
\bibitem [{\citenamefont {Shavitt}\ and\ \citenamefont
  {Bartlett}(2009)}]{shavitt_and_bartlett}%
  \BibitemOpen
  \bibfield  {author} {\bibinfo {author} {\bibfnamefont {I.}~\bibnamefont
  {Shavitt}}\ and\ \bibinfo {author} {\bibfnamefont {R.}~\bibnamefont
  {Bartlett}},\ }\href@noop {} {\emph {\bibinfo {title} {Many-Body Methods in
  Chemistry and Physics: MBPT and Coupled-Cluster Theory}}},\ Cambridge
  Molecular Science\ (\bibinfo  {publisher} {Cambridge University Press},\
  \bibinfo {year} {2009})\BibitemShut {NoStop}%
\bibitem [{\citenamefont {Crawford}\ and\ \citenamefont
  {Schaefer}(2007)}]{crawford_and_schaefer}%
  \BibitemOpen
  \bibfield  {author} {\bibinfo {author} {\bibfnamefont {T.~D.}\ \bibnamefont
  {Crawford}}\ and\ \bibinfo {author} {\bibfnamefont {H.~F.}\ \bibnamefont
  {Schaefer}},\ }\enquote {\bibinfo {title} {An introduction to coupled cluster
  theory for computational chemists},}\ in\ \href@noop {} {\emph {\bibinfo
  {booktitle} {Reviews in Computational Chemistry}}}\ (\bibinfo  {publisher}
  {John Wiley \& Sons, Inc.},\ \bibinfo {year} {2007})\ pp.\ \bibinfo {pages}
  {33--136}\BibitemShut {NoStop}%
\bibitem [{\citenamefont {{G. D. Purvis III}}\ and\ \citenamefont
  {Bartlett}(1982)}]{Purvis:82}%
  \BibitemOpen
  \bibfield  {author} {\bibinfo {author} {\bibnamefont {{G. D. Purvis III}}}\
  and\ \bibinfo {author} {\bibfnamefont {R.~J.}\ \bibnamefont {Bartlett}},\
  }\href {\doibase doi:10.1063/1.443164} {\bibfield  {journal} {\bibinfo
  {journal} {J. Chem. Phys.}\ }\textbf {\bibinfo {volume} {76}},\ \bibinfo
  {pages} {1910} (\bibinfo {year} {1982})}\BibitemShut {NoStop}%
\bibitem [{\citenamefont {Raghavachari}\ \emph {et~al.}(1989)\citenamefont
  {Raghavachari}, \citenamefont {Trucks}, \citenamefont {Pople},\ and\
  \citenamefont {Head-Gordon}}]{Raghavachari:89}%
  \BibitemOpen
  \bibfield  {author} {\bibinfo {author} {\bibfnamefont {K.}~\bibnamefont
  {Raghavachari}}, \bibinfo {author} {\bibfnamefont {G.~W.}\ \bibnamefont
  {Trucks}}, \bibinfo {author} {\bibfnamefont {J.~A.}\ \bibnamefont {Pople}}, \
  and\ \bibinfo {author} {\bibfnamefont {M.}~\bibnamefont {Head-Gordon}},\
  }\href {\doibase doi:10.1016/S0009-2614(89)87395-6} {\bibfield  {journal}
  {\bibinfo  {journal} {Chem. Phys. Lett.}\ }\textbf {\bibinfo {volume}
  {157}},\ \bibinfo {pages} {479} (\bibinfo {year} {1989})}\BibitemShut
  {NoStop}%
\bibitem [{\citenamefont {Mukherjee}\ \emph {et~al.}(1975)\citenamefont
  {Mukherjee}, \citenamefont {Moitra},\ and\ \citenamefont
  {Mukhopadhyay}}]{mukherjee_correlation_1975}%
  \BibitemOpen
  \bibfield  {author} {\bibinfo {author} {\bibfnamefont {D.}~\bibnamefont
  {Mukherjee}}, \bibinfo {author} {\bibfnamefont {R.~K.}\ \bibnamefont
  {Moitra}}, \ and\ \bibinfo {author} {\bibfnamefont {A.}~\bibnamefont
  {Mukhopadhyay}},\ }\href {\doibase 10.1080/00268977500103351} {\bibfield
  {journal} {\bibinfo  {journal} {Mol. Phys.}\ }\textbf {\bibinfo {volume}
  {30}},\ \bibinfo {pages} {1861} (\bibinfo {year} {1975})}\BibitemShut
  {NoStop}%
\bibitem [{\citenamefont {K\"{o}hn}\ \emph {et~al.}(2013)\citenamefont
  {K\"{o}hn}, \citenamefont {Hanauer}, \citenamefont {M\"{u}ck}, \citenamefont
  {Jagau},\ and\ \citenamefont {Gauss}}]{kohn_state-specific_2013}%
  \BibitemOpen
  \bibfield  {author} {\bibinfo {author} {\bibfnamefont {A.}~\bibnamefont
  {K\"{o}hn}}, \bibinfo {author} {\bibfnamefont {M.}~\bibnamefont {Hanauer}},
  \bibinfo {author} {\bibfnamefont {L.~A.}\ \bibnamefont {M\"{u}ck}}, \bibinfo
  {author} {\bibfnamefont {T.-C.}\ \bibnamefont {Jagau}}, \ and\ \bibinfo
  {author} {\bibfnamefont {J.}~\bibnamefont {Gauss}},\ }\href {\doibase
  10.1002/wcms.1120} {\bibfield  {journal} {\bibinfo  {journal} {Comp. Mol.
  Science}\ }\textbf {\bibinfo {volume} {3}},\ \bibinfo {pages} {176} (\bibinfo
  {year} {2013})}\BibitemShut {NoStop}%
\bibitem [{\citenamefont {Krylov}(2001)}]{krylov_size-consistent_2001}%
  \BibitemOpen
  \bibfield  {author} {\bibinfo {author} {\bibfnamefont {A.~I.}\ \bibnamefont
  {Krylov}},\ }\href {\doibase 10.1016/S0009-2614(01)00287-1} {\bibfield
  {journal} {\bibinfo  {journal} {Chem. Phys. Lett.}\ }\textbf {\bibinfo
  {volume} {338}},\ \bibinfo {pages} {375} (\bibinfo {year}
  {2001})}\BibitemShut {NoStop}%
\bibitem [{\citenamefont {Robinson}\ and\ \citenamefont
  {Knowles}(2011)}]{robinson_approximate_2011}%
  \BibitemOpen
  \bibfield  {author} {\bibinfo {author} {\bibfnamefont {J.~B.}\ \bibnamefont
  {Robinson}}\ and\ \bibinfo {author} {\bibfnamefont {P.~J.}\ \bibnamefont
  {Knowles}},\ }\href {\doibase 10.1063/1.3615060} {\bibfield  {journal}
  {\bibinfo  {journal} {J. Chem. Phys.}\ }\textbf {\bibinfo {volume} {135}},\
  \bibinfo {pages} {044113} (\bibinfo {year} {2011})}\BibitemShut {NoStop}%
\bibitem [{\citenamefont {Kowalski}\ and\ \citenamefont
  {Piecuch}(2000)}]{kowalski_method_2000}%
  \BibitemOpen
  \bibfield  {author} {\bibinfo {author} {\bibfnamefont {K.}~\bibnamefont
  {Kowalski}}\ and\ \bibinfo {author} {\bibfnamefont {P.}~\bibnamefont
  {Piecuch}},\ }\href {\doibase doi:10.1063/1.481769} {\bibfield  {journal}
  {\bibinfo  {journal} {J. Chem. Phys.}\ }\textbf {\bibinfo {volume} {113}},\
  \bibinfo {pages} {18} (\bibinfo {year} {2000})}\BibitemShut {NoStop}%
\bibitem [{\citenamefont {Small}\ and\ \citenamefont
  {Head-Gordon}(2012)}]{small:114103}%
  \BibitemOpen
  \bibfield  {author} {\bibinfo {author} {\bibfnamefont {D.~W.}\ \bibnamefont
  {Small}}\ and\ \bibinfo {author} {\bibfnamefont {M.}~\bibnamefont
  {Head-Gordon}},\ }\href {\doibase 10.1063/1.4751485} {\bibfield  {journal}
  {\bibinfo  {journal} {J. Chem. Phys.}\ }\textbf {\bibinfo {volume} {137}},\
  \bibinfo {eid} {114103} (\bibinfo {year} {2012})}\BibitemShut {NoStop}%
\bibitem [{\citenamefont {Meyer}(1971)}]{Meyer:71}%
  \BibitemOpen
  \bibfield  {author} {\bibinfo {author} {\bibfnamefont {W.}~\bibnamefont
  {Meyer}},\ }\href {\doibase doi:10.1002/qua.560050839} {\bibfield  {journal}
  {\bibinfo  {journal} {Int. J. Quantum Chem. Symp.}\ }\textbf {\bibinfo
  {volume} {5}},\ \bibinfo {pages} {341} (\bibinfo {year} {1971})}\BibitemShut
  {NoStop}%
\bibitem [{\citenamefont {Neese}\ \emph {et~al.}(2009)\citenamefont {Neese},
  \citenamefont {Wennmohs},\ and\ \citenamefont {Hansen}}]{Neese:09}%
  \BibitemOpen
  \bibfield  {author} {\bibinfo {author} {\bibfnamefont {F.}~\bibnamefont
  {Neese}}, \bibinfo {author} {\bibfnamefont {F.}~\bibnamefont {Wennmohs}}, \
  and\ \bibinfo {author} {\bibfnamefont {A.}~\bibnamefont {Hansen}},\ }\href
  {\doibase doi:10.1063/1.3086717} {\bibfield  {journal} {\bibinfo  {journal}
  {J. Chem. Phys.}\ }\textbf {\bibinfo {volume} {130}},\ \bibinfo {pages}
  {114108} (\bibinfo {year} {2009})}\BibitemShut {NoStop}%
\bibitem [{\citenamefont {Bartlett}\ and\ \citenamefont
  {{Musia\l}}(2006)}]{bartlett_addition_2006}%
  \BibitemOpen
  \bibfield  {author} {\bibinfo {author} {\bibfnamefont {R.~J.}\ \bibnamefont
  {Bartlett}}\ and\ \bibinfo {author} {\bibfnamefont {M.}~\bibnamefont
  {{Musia\l}}},\ }\href {\doibase 10.1063/1.2387952} {\bibfield  {journal}
  {\bibinfo  {journal} {J. Chem. Phys.}\ }\textbf {\bibinfo {volume} {125}},\
  \bibinfo {pages} {204105} (\bibinfo {year} {2006})}\BibitemShut {NoStop}%
\bibitem [{\citenamefont {Huntington}\ and\ \citenamefont
  {Nooijen}(2010)}]{huntington_pccsd:_2010}%
  \BibitemOpen
  \bibfield  {author} {\bibinfo {author} {\bibfnamefont {L.~M.~J.}\
  \bibnamefont {Huntington}}\ and\ \bibinfo {author} {\bibfnamefont
  {M.}~\bibnamefont {Nooijen}},\ }\href {\doibase doi:10.1063/1.3494113}
  {\bibfield  {journal} {\bibinfo  {journal} {J. Chem. Phys.}\ }\textbf
  {\bibinfo {volume} {133}},\ \bibinfo {pages} {184109} (\bibinfo {year}
  {2010})}\BibitemShut {NoStop}%
\bibitem [{\citenamefont {Mirman}(1973)}]{m73}%
  \BibitemOpen
  \bibfield  {author} {\bibinfo {author} {\bibfnamefont {R.}~\bibnamefont
  {Mirman}},\ }\href {\doibase 10.1007/BF02832643} {\bibfield  {journal}
  {\bibinfo  {journal} {Nuovo Cim. B}\ }\textbf {\bibinfo {volume} {18}},\
  \bibinfo {pages} {110} (\bibinfo {year} {1973})}\BibitemShut {NoStop}%
\bibitem [{\citenamefont {Leinaas}\ and\ \citenamefont {Myrheim}(1977)}]{lm77}%
  \BibitemOpen
  \bibfield  {author} {\bibinfo {author} {\bibfnamefont {J.~M.}\ \bibnamefont
  {Leinaas}}\ and\ \bibinfo {author} {\bibfnamefont {J.}~\bibnamefont
  {Myrheim}},\ }\href {\doibase 10.1007/BF02727953} {\bibfield  {journal}
  {\bibinfo  {journal} {Nuovo Cim. B}\ }\textbf {\bibinfo {volume} {37}},\
  \bibinfo {pages} {1} (\bibinfo {year} {1977})}\BibitemShut {NoStop}%
\bibitem [{\citenamefont {Nesbet}(1958)}]{nesbet_brueckners_1958}%
  \BibitemOpen
  \bibfield  {author} {\bibinfo {author} {\bibfnamefont {R.~K.}\ \bibnamefont
  {Nesbet}},\ }\href {\doibase 10.1103/PhysRev.109.1632} {\bibfield  {journal}
  {\bibinfo  {journal} {Phys. Rev.}\ }\textbf {\bibinfo {volume} {109}},\
  \bibinfo {pages} {1632} (\bibinfo {year} {1958})}\BibitemShut {NoStop}%
\bibitem [{\citenamefont {\v{C}\'{\i}\v{z}ek}\ and\ \citenamefont
  {Paldus}(1980)}]{cizek:80}%
  \BibitemOpen
  \bibfield  {author} {\bibinfo {author} {\bibfnamefont {J.}~\bibnamefont
  {\v{C}\'{\i}\v{z}ek}}\ and\ \bibinfo {author} {\bibfnamefont
  {J.}~\bibnamefont {Paldus}},\ }\href {\doibase
  doi:10.1088/0031-8949/21/3-4/006} {\bibfield  {journal} {\bibinfo  {journal}
  {Physica Scripta}\ }\textbf {\bibinfo {volume} {21}},\ \bibinfo {pages} {251}
  (\bibinfo {year} {1980})}\BibitemShut {NoStop}%
\bibitem [{\citenamefont {Bartlett}\ and\ \citenamefont
  {Purvis}(1980)}]{bartlett:80}%
  \BibitemOpen
  \bibfield  {author} {\bibinfo {author} {\bibfnamefont {R.~J.}\ \bibnamefont
  {Bartlett}}\ and\ \bibinfo {author} {\bibfnamefont {G.~D.}\ \bibnamefont
  {Purvis}},\ }\href {\doibase doi:10.1088/0031-8949/21/3-4/007} {\bibfield
  {journal} {\bibinfo  {journal} {Physica Scripta}\ }\textbf {\bibinfo {volume}
  {21}},\ \bibinfo {pages} {255} (\bibinfo {year} {1980})}\BibitemShut
  {NoStop}%
\bibitem [{\citenamefont {Hampel}\ \emph {et~al.}(1992)\citenamefont {Hampel},
  \citenamefont {Peterson},\ and\ \citenamefont
  {Werner}}]{hampel_comparison_1992}%
  \BibitemOpen
  \bibfield  {author} {\bibinfo {author} {\bibfnamefont {C.}~\bibnamefont
  {Hampel}}, \bibinfo {author} {\bibfnamefont {K.~A.}\ \bibnamefont
  {Peterson}}, \ and\ \bibinfo {author} {\bibfnamefont {H.-J.}\ \bibnamefont
  {Werner}},\ }\href {\doibase 10.1016/0009-2614(92)86093-W} {\bibfield
  {journal} {\bibinfo  {journal} {Chem. Phys. Lett.}\ }\textbf {\bibinfo
  {volume} {190}},\ \bibinfo {pages} {1} (\bibinfo {year} {1992})}\BibitemShut
  {NoStop}%
\bibitem [{\citenamefont {Werner}\ \emph
  {et~al.}(2012{\natexlab{a}})\citenamefont {Werner}, \citenamefont {Knowles},
  \citenamefont {Knizia}, \citenamefont {Manby}, \citenamefont {{Sch\"{u}tz}}
  \emph {et~al.}}]{MOLPRO_brief}%
  \BibitemOpen
  \bibfield  {author} {\bibinfo {author} {\bibfnamefont {H.-J.}\ \bibnamefont
  {Werner}}, \bibinfo {author} {\bibfnamefont {P.~J.}\ \bibnamefont {Knowles}},
  \bibinfo {author} {\bibfnamefont {G.}~\bibnamefont {Knizia}}, \bibinfo
  {author} {\bibfnamefont {F.~R.}\ \bibnamefont {Manby}}, \bibinfo {author}
  {\bibfnamefont {M.}~\bibnamefont {{Sch\"{u}tz}}},  \emph {et~al.},\
  }\href@noop {} {\enquote {\bibinfo {title} {Molpro, version 2012.1, a package
  of ab initio programs},}\ } (\bibinfo {year} {2012}{\natexlab{a}}),\ \bibinfo
  {note} {see http://www.molpro.net}\BibitemShut {NoStop}%
\bibitem [{\citenamefont {Werner}\ \emph
  {et~al.}(2012{\natexlab{b}})\citenamefont {Werner}, \citenamefont {Knowles},
  \citenamefont {Knizia}, \citenamefont {Manby},\ and\ \citenamefont
  {Sch{\"u}tz}}]{MOLPRO-WIREs}%
  \BibitemOpen
  \bibfield  {author} {\bibinfo {author} {\bibfnamefont {H.-J.}\ \bibnamefont
  {Werner}}, \bibinfo {author} {\bibfnamefont {P.~J.}\ \bibnamefont {Knowles}},
  \bibinfo {author} {\bibfnamefont {G.}~\bibnamefont {Knizia}}, \bibinfo
  {author} {\bibfnamefont {F.~R.}\ \bibnamefont {Manby}}, \ and\ \bibinfo
  {author} {\bibfnamefont {M.}~\bibnamefont {Sch{\"u}tz}},\ }\href@noop {}
  {\bibfield  {journal} {\bibinfo  {journal} {WIREs Comput Mol Sci}\ }\textbf
  {\bibinfo {volume} {2}},\ \bibinfo {pages} {242} (\bibinfo {year}
  {2012}{\natexlab{b}})}\BibitemShut {NoStop}%
\bibitem [{\citenamefont {Werner}\ and\ \citenamefont {Knowles}(1988)}]{WK88}%
  \BibitemOpen
  \bibfield  {author} {\bibinfo {author} {\bibfnamefont {H.-J.}\ \bibnamefont
  {Werner}}\ and\ \bibinfo {author} {\bibfnamefont {P.~J.}\ \bibnamefont
  {Knowles}},\ }\href {\doibase 10.1063/1.455556} {\bibfield  {journal}
  {\bibinfo  {journal} {J. Chem. Phys.}\ }\textbf {\bibinfo {volume} {89}},\
  \bibinfo {pages} {5803} (\bibinfo {year} {1988})}\BibitemShut {NoStop}%
\bibitem [{\citenamefont {Knizia}\ \emph {et~al.}(2009)\citenamefont {Knizia},
  \citenamefont {Adler},\ and\ \citenamefont
  {Werner}}]{knizia_simplified_2009}%
  \BibitemOpen
  \bibfield  {author} {\bibinfo {author} {\bibfnamefont {G.}~\bibnamefont
  {Knizia}}, \bibinfo {author} {\bibfnamefont {T.~B.}\ \bibnamefont {Adler}}, \
  and\ \bibinfo {author} {\bibfnamefont {H.-J.}\ \bibnamefont {Werner}},\
  }\href {\doibase doi:10.1063/1.3054300} {\bibfield  {journal} {\bibinfo
  {journal} {J. Chem. Phys.}\ }\textbf {\bibinfo {volume} {130}},\ \bibinfo
  {pages} {054104} (\bibinfo {year} {2009})}\BibitemShut {NoStop}%
\bibitem [{\citenamefont {Van~Voorhis}\ and\ \citenamefont
  {Head-Gordon}(2000)}]{van_voorhis_benchmark_2000}%
  \BibitemOpen
  \bibfield  {author} {\bibinfo {author} {\bibfnamefont {T.}~\bibnamefont
  {Van~Voorhis}}\ and\ \bibinfo {author} {\bibfnamefont {M.}~\bibnamefont
  {Head-Gordon}},\ }\href {\doibase doi:10.1063/1.1319643} {\bibfield
  {journal} {\bibinfo  {journal} {J. Chem. Phys.}\ }\textbf {\bibinfo {volume}
  {113}},\ \bibinfo {pages} {8873} (\bibinfo {year} {2000})}\BibitemShut
  {NoStop}%
\bibitem [{\citenamefont {Hachmann}\ \emph {et~al.}(2006)\citenamefont
  {Hachmann}, \citenamefont {Cardoen},\ and\ \citenamefont
  {Chan}}]{hachmann_multireference_2006}%
  \BibitemOpen
  \bibfield  {author} {\bibinfo {author} {\bibfnamefont {J.}~\bibnamefont
  {Hachmann}}, \bibinfo {author} {\bibfnamefont {W.}~\bibnamefont {Cardoen}}, \
  and\ \bibinfo {author} {\bibfnamefont {G.~K.-L.}\ \bibnamefont {Chan}},\
  }\href {\doibase doi:10.1063/1.2345196} {\bibfield  {journal} {\bibinfo
  {journal} {J. Chem. Phys.}\ }\textbf {\bibinfo {volume} {125}},\ \bibinfo
  {pages} {144101} (\bibinfo {year} {2006})}\BibitemShut {NoStop}%
\bibitem [{\citenamefont {White}(1992)}]{white_density_1992}%
  \BibitemOpen
  \bibfield  {author} {\bibinfo {author} {\bibfnamefont {S.~R.}\ \bibnamefont
  {White}},\ }\href {\doibase 10.1103/PhysRevLett.69.2863} {\bibfield
  {journal} {\bibinfo  {journal} {Phys. Rev. Lett.}\ }\textbf {\bibinfo
  {volume} {69}},\ \bibinfo {pages} {2863} (\bibinfo {year}
  {1992})}\BibitemShut {NoStop}%
\bibitem [{\citenamefont {Chan}\ and\ \citenamefont
  {Head-Gordon}(2002)}]{chan_highly_2002}%
  \BibitemOpen
  \bibfield  {author} {\bibinfo {author} {\bibfnamefont {G.~K.-L.}\
  \bibnamefont {Chan}}\ and\ \bibinfo {author} {\bibfnamefont {M.}~\bibnamefont
  {Head-Gordon}},\ }\href {\doibase doi:10.1063/1.1449459} {\bibfield
  {journal} {\bibinfo  {journal} {J. Chem. Phys.}\ }\textbf {\bibinfo {volume}
  {116}},\ \bibinfo {pages} {4462} (\bibinfo {year} {2002})}\BibitemShut
  {NoStop}%
\bibitem [{\citenamefont {Chan}\ and\ \citenamefont
  {Sharma}(2011)}]{chan_density_2011}%
  \BibitemOpen
  \bibfield  {author} {\bibinfo {author} {\bibfnamefont {G.~K.-L.}\
  \bibnamefont {Chan}}\ and\ \bibinfo {author} {\bibfnamefont {S.}~\bibnamefont
  {Sharma}},\ }\href {\doibase 10.1146/annurev-physchem-032210-103338}
  {\bibfield  {journal} {\bibinfo  {journal} {Annu. Rev. Phys. Chem.}\ }\textbf
  {\bibinfo {volume} {62}},\ \bibinfo {pages} {465} (\bibinfo {year}
  {2011})}\BibitemShut {NoStop}%
\bibitem [{\citenamefont {Tsuchimochi}\ and\ \citenamefont
  {Scuseria}(2009)}]{tsuchimochi_strong_2009}%
  \BibitemOpen
  \bibfield  {author} {\bibinfo {author} {\bibfnamefont {T.}~\bibnamefont
  {Tsuchimochi}}\ and\ \bibinfo {author} {\bibfnamefont {G.~E.}\ \bibnamefont
  {Scuseria}},\ }\href {\doibase doi:10.1063/1.3237029} {\bibfield  {journal}
  {\bibinfo  {journal} {J. Chem. Phys.}\ }\textbf {\bibinfo {volume} {131}},\
  \bibinfo {pages} {121102} (\bibinfo {year} {2009})}\BibitemShut {NoStop}%
\bibitem [{\citenamefont {Sinitskiy}\ \emph {et~al.}(2010)\citenamefont
  {Sinitskiy}, \citenamefont {Greenman},\ and\ \citenamefont
  {Mazziotti}}]{sinitskiy_strong_2010}%
  \BibitemOpen
  \bibfield  {author} {\bibinfo {author} {\bibfnamefont {A.~V.}\ \bibnamefont
  {Sinitskiy}}, \bibinfo {author} {\bibfnamefont {L.}~\bibnamefont {Greenman}},
  \ and\ \bibinfo {author} {\bibfnamefont {D.~A.}\ \bibnamefont {Mazziotti}},\
  }\href {\doibase doi:10.1063/1.3459059} {\bibfield  {journal} {\bibinfo
  {journal} {J. Chem. Phys.}\ }\textbf {\bibinfo {volume} {133}},\ \bibinfo
  {pages} {014104} (\bibinfo {year} {2010})}\BibitemShut {NoStop}%
\end{thebibliography}

%

\end{document}